\magnification=\magstep 1
\raggedbottom
\tolerance=1000
\baselineskip 20 truept
\parindent 36 truept
\hsize 6.5 truein
\vsize 9 truein

\def\alt{\;{\lower 1.5pt\hbox{$<$} \above 0pt \raise 1.5pt\hbox{$\sim$}}\;}
\def\agt{\;{\lower 1.5pt\hbox{$>$} \above 0pt \raise 1.5pt\hbox{$\sim$}}\;}

\leftline{\bf Existence of algebraic decay in non-Abelian ferromagnets}
\smallskip
\leftline{\indent A. Patrascioiu}
\leftline{\indent \it Physics Department and Center for the Study of Complex Systems}
\leftline{\indent \it University of Arizona, Tucson, AZ 85721}
\smallskip
\leftline{\indent (Received December 1, 1991)}
\vskip 1pc
\vbox{

\hsize=6truein
\item{} \ \ \ \ \ The low temperature regime of non-Abelian two dimensional ferromagnets
is investigated. The method involves mapping such models into certain site-bond
percolation processes and using ergodicity in a novel fashion. It is concluded
that all ferromagnets possessing a continuous symmetry (Abelian or not) exhibit
algebraic decay of correlations at sufficiently low temperatures.
}
\smallskip
\line{\indent PACS: 05.07.Fh.05.50+q.11.15Ha.64.60t \hfil}
\eject 

In a recent letter [1] Seiler and I proposed studying the phase structure
of the 2D $O(N)$ models by mapping them into a correlated site-bond percolation 
problem. This approach was applied to certain discrete spin modes and to the $O(2)$ model, 
for which we rederived the Froehlich and Spencer [2] result regarding the existence of a
massless phase at sufficiently low temperatures $1/\beta$. In
this paper I report an extension of the percolation approach to $O(N)$ $N \geq 3$. It leads to the conclusion that a massless phase exists in all $O(N)$ models.

For completeness I will repeat the main points of Ref.~[1] (see also
Ref.~[3] for a more complete discussion). With any $O(N)$ spin configuration one can
associate an Ising spin configuration by dividing the sphere $S(N-1)$ into two hemispheres 
and introducing an Ising variable $\sigma = \pm 1$, which specifies in which
hemisphere the spin points. In this manner the standard nearest neighbor action (s.n.n.a.) 
for the $O(N)$ model allows rewriting the partition function as
$$Z = \sum_{\{\sigma\}} \left(\prod_{i\in\Lambda} \int d s_{\parallel i} d\vec s_{pi}\right) 
\cdot \exp \left[\beta \sum_{\langle i,j\rangle} (s_{\parallel i} s_{\parallel j} \sigma_i
\sigma_j + \vec s_{pi} \cdot \vec s_{pj})\right] \eqno(1)$$

\noindent Here $\vec u$ is the unit vector chosen for specifying the hemispherical 
decomposition, $s_\parallel=|\vec s\cdot\vec u|$ and $\vec s_p\cdot\vec u=0$. 
With respect to the Ising variables the action is ferromagnetic, hence amenable to 
the Fortuin-Kasteleyn transformation [4].  This 
procedure associates to the Ising problem a correlated site-bond percolation
process defined as follows: 
\item{} FK1-identify clusters of like-$\sigma$ spins (H-clusters)
\item{} FK2-within each H-cluster occupy bonds randomly with probability $1-exp(-2
\beta_{ij})$ (obtain FK-cluster)
\item{} FK3-assign to every site within a given FK-cluster the same $\sigma$ value, 
obtained by choosing randomly + or - with probability 1/2.

\noindent Here $\beta_{ij}$ is the space dependent inverse temperature, which for the 
s.n.n.a. would be $\beta s_{\parallel i}s_{\parallel j}$. Fortuin and Kasteleyn proved 
that the mean FK-cluster size (expected size of the cluster attached to the origin) 
equals the magnetic susceptibility of the Ising variable
$$\chi_{_{Is}} \equiv {1\over |\Lambda|} \sum_{x,y\in \Lambda} \langle\sigma_x \sigma_y
\rangle\ .\eqno(2)$$

\noindent In particular the latter diverges when the mean FK-cluster size diverges.

To apply the F-K procedure to the $O(N)$ models, Seiler and I considered
a modified model called `cut' action: the Gibbs factor is s.n.n.a. only if $|\vec s_i - \vec s_j|<\epsilon, \ 0< \epsilon <2$ and 0 otherwise. We then formulated the following three conjectures:

\noindent C1: The Mermin-Wagner theorem applies to the `cut' model.

\noindent C2: The $O(N)$ models (`cut' or not) are ergodic.

\noindent C3: On a triangular lattice $T$ a percolation process produced by a measure enjoying the symmetries of the lattice can contain at most one percolating cluster. 

\noindent I refer the reader to Refs.~[1] and [3] for a thorough discussion of the motivations
behind these three conjectures and of the comparison of the `cut' and the s.n.n.a. 
models.  I will elaborate only on C2, which is central
to the arguments presented in this paper. Imagine a very large lattice on which one 
has used the Monte Carlo procedure to simulate the $O(N)$ model. If one has achieved 
thermalization, then this configuration is `typical.' In the infinite 
volume limit a typical configuration has two important properties:
\item{} P1: spacial averages equal ensemble averages (Birkoff's theorem)
\item{} P2: the configuration is (statistically) invariant under additional Monte Carlo steps.

I will briefly sketch the argument used in Ref.~[1] to prove that the `cut' $O(2)$ 
model must exhibit algebraic decay of its correlation functions for $\epsilon$ sufficiently small. 
In Eq.~(2) let $\langle\cdot\rangle$ stand for expectation value measured
with the full Gibbs measure. By P1 and P2, $\chi_{_{Is}}$ can be computed as a quenched 
expectation value provided the spins $s_{i||}$ are assigned the values of a typical configuration. Since the Gibbs measure is invariant under lattice translations 
and (discrete) rotations, by C1 and C3 a typical configuration cannot contain a percolating H-cluster. An interesting theorem by Russo [5] states that if a
translational invariant percolation process on a $T$ lattice is such that neither clusters of the set $E$ nor of its complement $\bar E$ percolate, then the mean 
cluster size of both $E$ and $\bar E$ must diverge. Taking $E$ to stand for $\sigma=+1$ and
$\bar E$ for $\sigma=-1$ shows that the mean size of the H-clusters must diverge. (This statement is not surprising since at $\beta=0$ and $\epsilon=2$ the $O(N)$  model 
is equivalent to the Bernoulli site-percolation process with $p=1/2$ and for the  latter 
the critical density on a $T$ lattice is indeed 1/2.) The FK-clusters are subclusters of the H-clusters obtained via rule FK2. In the `cut' model, this rule must be 
amended. Indeed because of rule FK3, the constraint could be violated
unless bonds are occupied at all sites having $s_\parallel > d \equiv \epsilon/2$. 
Therefore, in a `cut' $O(N)$ model, the FK-clusters must contain D-clusters defined by 
the condition $s_\parallel>d$. In Ref.~[1] we showed that for the `cut' $O(2)$ model simple 
applications of C1 and C3 required that neither $D$-clusters nor 
$\bar D$-clusters	
 $(s_{||}<d)$ can percolate and then, by Russo's theorem, both must have
divergent mean size, QED.
\smallskip
\indent
From the discussion presented thus far it follows that in any `cut' $O(N)$ $N>1$ model on a $T$ lattice, if neither clusters of $D$ nor of $\bar D$ percolate,
the mean FK-cluster size must diverge and hence correlations must decay algebraically. In fact in the `cut' model $D$-clusters can never percolate. Indeed the set $D$ 
consists of two disconnected pieces, both of which are contained in H-clusters and I
have already argued that H-clusters cannot percolate. Therefore, the only question is whether $\bar D$-clusters could percolate for $\epsilon$ sufficiently small?
The reason for which a topological answer to this question exists in $O(2)$ is that in that case the set $\bar D$ consists also of two disconnected pieces, which, for 
$\epsilon<\sqrt 2$, cannot communicate. Obviously in $O(N) N \geq 3$, $\bar D$ is a connected set and a new strategy must be employed. In the sequel I will state three independent
arguments,  that in the `cut' $O(N)$ model $\bar D$-clusters cannot percolate for $\epsilon$ sufficiently small or $\beta$ sufficiently large. Each argument requires a  new 
conjecture and I will address their merits too.
\eject
\leftline{\bf Argument 1}
\smallskip
This is a proof by contradiction. For simplicity I will discuss the 
s.n.n.a. $O(3)$ model $(\epsilon=2)$ at $\beta$ large and choose 
$\vec u=\hat z$. I will take $d$ small but independent of $\beta$ 
- so that by FK2, when $\beta$ is large,the bond occupation
probability for sites in $D$ goes to 1. I will assume that a cluster 
of $\bar D$  
percolates and show that that assumption suggests that a certain 
magnetic susceptibility (Eq.~(4)) diverges. To that end I introduce 
spherical coordinates  and rewrite the partition function as
$$Z = \left(\prod_{i\in \Lambda} \int^\pi_0 d \theta_i \int^{2\pi}_0 
d\varphi_i \right) \cdot \exp\left\{\beta \sum_{\langle i,j\rangle} 
\left[\cos\theta_i\cos\theta_j + \sin\theta_i\sin\theta_j\cos(\varphi_i 
- \varphi_j\right]\right\}\eqno(3)$$
\noindent Consider the following susceptibility
$$\chi_\varphi \equiv {1\over |\Lambda|} \sum_{x_i y 
\in \Lambda} \langle\cos(\varphi_x-\varphi_y)\rangle\eqno(4)$$

\noindent By P1 and P2 $\chi_{\varphi}$ could be measured by 
quenching the $\theta$ variables to the values $\bar \theta$ 
they would take in a typical configuration. That is 
$$\chi_\varphi = {1\over |\Lambda|} \sum_{x_i y \in \Lambda} 
\langle\cos(\varphi_x-\varphi_y)\rangle_q \eqno(5)$$

\noindent where $\langle\cdot\rangle_q$ means expectation value 
computed with the measure
$$\left(\prod_{i\in\Lambda} \int^{2\pi}_0 d \varphi_i\right) \exp
\left[\beta \sum_{\langle i,j\rangle} \sin\bar\theta_i\sin
\bar\theta_j \cdot \cos(\varphi_i-\varphi_j)\right]\ .\eqno(6)$$

\noindent Since the quenched model is an $O(2)$ model (albeit with 
space dependent 
couplings), one can employ Ginibre's inequality [6] to bound 
$\chi_{\varphi}$ from below by the value it would take if in 
the measure (6) one replaced 
$\beta\sin\theta_i\sin\theta_j$ by 0 at all sites where 
$\sqrt{\beta\sin\theta_i}<c$ for some  $c>0$. Under the 
assumption that $\bar D$ percolates, by C3, these sites could not
possibly percolate, but would form islands. The average size of these 
islands relative to the average distance between them would decrease 
with beta. Indeed by the Mermin-Wagner theorem, the probability of 
finding the spin at a site taking 
values in some subset of the sphere $A$ of volume $V(A)$ is equal 
to $V(A)/4\pi$.  (For $\beta$ large, one can use perturbation 
theory to estimate the average size of  
these islands, which becomes actually independent of $\beta$.) 
Thus the assumption that the equatorial strip $\bar D$ percolates 
implies that $\chi_{\varphi}$ is bounded from below
by the susceptibility of an $O(2)$ model at large inverse 
temperature, but on a lattice having some small, randomly distributed 
holes. Although I am not aware of any rigorous result proving that, 
the following conjecture seems eminently reasonable.
\medskip
\noindent C4: Consider a $T$ lattice and dilute bonds randomly 
with a probability smaller than the percolation probability for 
unoccupied bonds. Then there exists a $\beta_{kt}<\infty$ such 
that for any $\beta>\beta_{kt}$ the susceptibility diverges.

Before motivating this conjecture, let me say that there is no 
reason to 
expect that if in the $O(3)$ model $\bar D$ percolated, the 
polar caps would be distributed as the holes produced by a 
Bernoulli process. Their actual distribution
would be controlled by the full $O(3)$ measure. However, if 
$\bar D$ percolated and especially if the model had a mass 
gap, by some central limit theorem, one
would expect the polar caps to form islands and their distribution 
to be random at distances much larger than the correlation length.

The intuition for C4 comes from the following rigorous results:
\item{} a) Georgii [7] proved that if one randomly dilutes sites or 
bonds on a regular lattice with $D \geq 2$, then provided a remaining 
cluster percolates, there exists an inverse temperature $\beta_c<\infty$ such that for $\beta>\beta_c$ there exists long range 
order (l.r.o.).
\item{} b) De Massi {\it et al.} [8], proved that under the same conditions 
as above, the Laplacian retains its continuous spectrum.

\noindent In the language of the Coulomb gas, my conjecture is that 
if one introduces in the gas perfect conductors, randomly distributed, 
if the perfectly conducting
regions do not percolate, at sufficiently low temperatures, the 
Coulomb gas does not exhibit Debye screening (the introduction of the perfect conductors
will only affect the dielectric constant).

To conclude this argument, the contradiction is this: if one assumes 
that for the $O(3)$ model $\bar D$ percolates and $\chi_{_{Is}}$ is finite, 
then clearly so is the $s_z$-susceptibility (since $s_z\leq 1$). 
On the other hand C4 strongly suggests that
the $s_x-s_y$ susceptibility would diverge when $\beta$ is large. 
This is a clear violation of $O(3)$ invariance, hence the assumption 
that $\bar D$ percolates must be false. Although not transparent, 
the topology of $O(3)$ is crucial for this argument.  Indeed one may 
wonder if a similar reasoning could not be used to relate 
the $O(2)$ model to the Ising model and thus prove that the latter 
must exhibit l.r.o. at large $\beta$, in violation of the 
Mermin-Wagner theorem? The answer is no, precisely because $\bar D$ 
is no longer a connected set and thus it could not possibly percolate.
\bigskip
\leftline{\bf Argument 2}
\smallskip
This is again a proof by contradiction. For simplicity I consider the 
`cut' $O(3)$ model and choose $\vec u=\hat z$. I would like to argue 
that if the equatorial strip $\bar D$ percolated, then $O(3)$ 
invariance would be broken. 

Next let me consider the realistic case of a $T$ lattice and an 
$\epsilon$ small,
yet $\epsilon>0$. Suppose that in fact the equatorial strip 
$\bar D$ does percolate and hence its complement $D$ forms islands.  
In the `cut' model, the lines $s_z=c>d$ will have
to form closed loops, nested inside these islands. Consider now a 
c-tilted equator, namely the great circle passing thru $s_z=c$ and 
$s_x=0$. Since neither the hemisphere $s_x>0$ nor $s_x<0$ can 
percolate, any site of the lattice must be 
surrounded by an infinite sequence $X(k)\ k\in Z$ of concentric 
closed loops $s_x=0$. (By the line $s_x=0$ I mean a line on the 
dual lattice such that $s_{x_i}\cdot s_{x_j}\leq 0$; same type 
of qualifications apply to all other lines appearing in this discussion.) 
$O(3)$ invariance requires that the average number of intersections 
of the $X$ lines with the c-tilted equators is independent of c.  
However if $\bar D$ percolates, then in any typical configuration 
there exists a $k_0<\infty$ such that any $X(k)$ line with
$k>k_0$ intersects the percolating cluster. That means that infinitely 
many $X$ lines cross the $c=0$ tilted equator, while they may or may not 
cross the c-tilted equators with $c>0$. In other words if $\bar D$ 
percolates, then one would expect the average number of crossings of 
the $X$ lines with the c-tilted equators to decrease with c, in violation of $O(3)$ invariance. If on the other hand $\bar D$ does not percolate, 
then both $D$ and $\bar D$ form rings and no a priori asymmetry in
the average number of crossings of the $X$ lines with the c-tilted equators
exists. (An example where $\bar D$ percolates is the Richard model [9],  
which is a  modified $O(3)$ model in which $|s_z|<1-b$ for some 
$b>0$, hence this model is only  $O(2)$ invariant. The percolation 
approach used in Ref.~[1] can be employed to prove rigorously that 
this model has to be massless for $\epsilon$ sufficiently small - see Ref.~[3]; $\chi_{_{\varphi}}$ diverges, yet $\chi_{_{Is}} < \infty$.) 
 
In the discussion above I used the word `expect' because one could say 
that even though if $\bar D$ percolates the regions with $s_z>c$ 
are hidden inside regions of smaller $s_z$ values, they are larger 
and thus restore $O(3)$ invariance. However $O(3)$ invariance 
requires that any typical configuration has the following two properties:   
\item{} T1: The area is preserved.
\item{} T2: The gradient is preserved.

\noindent Property T1 means that the density of sites where the 
spin points in some region
A is proportional to the volume $V(A)$. Property T2 says that if 
one selects two points on the sphere $p_1$ and $p_2$, separated 
by a distance $L$, the average distance 
between sites where the spin points in the neighborhood of $p_1$ 
respectively $p_2$ depends only on $L$ (it is independent of which 
$p_1$ and $p_2$ are chosen, provided  they are at distance $L$). 
Obviously both properties are required by C1.
\medskip
\noindent C5: If in the `cut' $O(3)$ model $\bar D$ percolated, 
then the typical configuration would violate $T_1$ or $T_2$ (or 
both) and, hence, $O(3)$ invariance.

The motivation for C5 is this: if $\bar D$ percolated, then, 
as already argued, $D$ would form islands - as opposed to rings, 
which are formed when neither $\bar D$ nor $D$ percolates on a 
$T$ lattice. The basic difference between a system forming islands 
and one forming rings is that islands are basically of finite
size - the probability to find an island of diameter $L$ decreases 
exponentially
with $L$; on the contrary, if the system forms rings, there exists 
an infinite sequence of clusters surrounding each other and hence 
no exponential suppression 
of large clusters. Thus if the system forms islands the typical 
configuration will contain mostly mappings of a hemisphere over some 
finite region of $T$. It is easy to check that such maps cannot 
preserve both T1 and T2. No such difficulty exists if one considers 
rings - arbitrarily large regions of $T$. 
\bigskip
\leftline{\bf Argument 3}
\smallskip
As I have already noted, if $\bar D$ percolates, then $D$ forms 
islands. Moreover, $D$ consists of two disconnected pieces $D_u$ 
and $D_l$. When $\epsilon$ is sufficiently small, the volume of 
$D_u$, $V(D_u)$ is much larger than that of $\bar D$, $V(\bar D)$.  
On the other hand the area of the boundary of $D_u$, $S(D_u)$ is 
half $S(\bar D)$. Is it reasonable to expect that under these 
circumstances, the mean cluster size of $D_u$ is finite while 
that of $\bar D$ infinite? The answer is provided by the following 
conjecture:
\medskip
\noindent C6: In the `cut' $O(N)$ model, if two sets $A$ and 
$B$ have V(A)=V(B) and $S(A)<S(B)$, then there exists $\epsilon_0 
(A,B)>0$ such that for any $\epsilon<\epsilon_0,$ 
$\langle A\rangle \geq \langle B\rangle$, where $\langle\cdot
\rangle$ represents the mean cluster size.

The conjecture says that at given volume, the larger the 
surface of a
set, the smaller its average cluster size. The reason for adding 
the qualifier that $\epsilon < \epsilon_0$ is that for 
$\epsilon>0$ the surface of the clusters of a set 
A need not consist of points on the surface of A. I believe that 
this conjecture  is intuitively clear. It can be proved in 1D. 
In 2D it was verified numerically for $O(3)$ as follows: A was 
the Northern polar cap of area $4\pi/3$, B the equatorial strip 
of the same area and $\epsilon$ was such that the Northern and 
Southern polar caps  could barely communicate. The data indicated 
that the mean cluster size of both A and B increased as $L^{2-\eta}$ 
($L$-linear size of the lattice) and that $\eta_A<\eta_B$.

If C6 is true, it cannot be true that in the `cut' $O(N)$ 
models the equatorial strip $\bar D$ percolates. Indeed 
if $\bar D$ percolates, its mean cluster size is divergent. 
By C6, for $\epsilon$ sufficiently small, so is the mean 
cluster size of $D$. By Russo's theorem, on a $T$ lattice, 
that can occur only if neither $D$ nor $\bar D$ percolates. 
QED.  
\bigskip
\eject
\leftline{\bf Discussion}
\smallskip
The arguments presented above indicate that all 2D $O(N)$ models possess
a massless phase.  (This situation contradicts common wisdom.  Evidence
in favor of the latter is analyzed separately [10] and found wanting.)
The arguments moreover suggest that although at large $\beta$ extended
topological defects - instantons - may exist in non-Abelian models, they
are supressed entropically with respect to spin waves.  This situation,
already conjectured by the author in 1986 [11], suggests that for
 $$N \geq 3$$ the 2-point function may behave as
$$\langle\vec s_\circ \cdot \vec s_x\rangle \sim a(\beta) {e^{-m(\beta)x}
\over \sqrt x} + b(\beta) {1\over x^{\eta(\beta)}}\ .\eqno(7)$$
\noindent I have no basis at the present time to estimate $a(\beta)$ 
and $b(\beta)$, nor whether $\eta$ depends on $\beta$ in any given model.  However, it could be that $a$ and $b$ are such that at intermediate distances the decay is exponential to a very good approximation (a similar effect governs the time evolution of a 
metastable state in nonrelativistic quantum mechanics [12]).

Finally a word about perturbation theory. The fact that the 2D 
$O(N)$ models possess a massless phase for $\beta$ sufficiently 
large does not imply that in 2D
perturbation theory fails to produce the correct asymptotic 
expansion at fixed distances (as it does in 1D for $N\geq 3$). 
However if one defines the Callan-Symanzik $\beta$-function 
by requiring that say $\langle\vec s(0)\cdot\vec s(x)\rangle/\langle 
\vec s(0)\cdot\vec s(y)\rangle$ is a
renormalization group invariant for $x,y\gg 1$, then clearly 
an algebraic decay for
$\beta >\beta_{kt}(N)$ implies that the Callan-Symanzik 
$\beta$-function could be chosen to be vanishing. If my conjecture 
about Eq.~(7) proved to be correct, one could also define the 
$\beta$-function as $d\beta/dln(m)$, in which case one may
find the famous asymptotic freedom answer. However I find it 
hard to believe that if that were the case, the continuum limit 
constructed by letting  $\beta \to \infty$ would not contain 
(coupled) massless excitations (of course a
continuum limit could also be constructed for any $\infty>\beta>
\beta_{kt}(N)$ - that field theory would be a massless theory).

Many of the ideas expressed in this paper stem from my long time collaboration
with Erhard Seiler.  I am also gratefulfor the hospitality extended to me by the
Max Planck Institut fur Pysik und Astrophysik - Munich.
\eject
\leftline{\bf References}
\medskip
\item{[1]} A.~Patrascioiu and E.~Seiler,  
Phys. Rev. Lett. {\bf 68}, 1395 (1992).
\item{[2]} J.~Froehlich and T.~Spencer, Comm.~Math.~Phys.~{\bf 81}, 455 (1981).
\item{[3]} A.~Patrascioiu and E.~Seiler, J.Stat.Phys. {\bf 69}, 55 (1992). 
\item{[4]} C.~M.~Fortuin and P.~W.~Kasteleyn, J.~Phys.~Soc.~JPN (suppl.) {\bf 24}, 86 (1969).
\item{[5]} L.~Russo, Z.~Wahrsch. Verw. Gebiete {\bf 42}, 39 (1978).
\item{[6]} J.~Ginibre, Comm.~Math.~Phys.~{\bf 16}, 310 (1970).
\item{[7]} H.~O.~Georgii, Comm.~Math.~Phys.~{\bf 81}, 527 (1981).
\item{[8]} A.~DeMassi,P.~A.~Ferrari, S.~Goldstein and W.~D.~Wick, 
J.~Stat.~Phys.~{\bf 55}, 787 (1989).
\item{[9]} J.-L.~Richard, Phys.~Lett.~B {\bf 134}, 75 (1987).
\item{[10]} A. Patrascioiu and E. Seiler, The Difference between Abelian
and Non-Abelian Models: Facts and Fancy, MPI preprint, 1991, math-ph/9903038.
\item{[11]} A. Patrascioiu, Phys. Rev. Lett. {\bf 58}, 2285 (1987).
\item{[12]} A. Patrascioiu, Phys. Rev. D {\bf 24}, 496 (1981).

\end
The arguments presented above pertain mostly to the `cut' $O(N)$ 
models. This modification of the s.n.n.a. models does not change 
their perturbative behavior
nor does it eliminate instantons in those cases where they exist. 
For many years now, there has been total unanimity among condensed 
matter and particle physicists that there is a fundamental difference 
between Abelian and non-Abelian $\sigma$ models in 2D and gauge 
theories in 4D. It is claimed that the difference stems from either 
the instantons or the asymptotic freedom present in non-Abelian models, 
but not in Abelian ones. That such reasoning is rather glib can 
easily be seen. For example among $O(N)$ models only $O(3)$ has 
instantons. Also Richard's 
model is asymptotically free, yet massless (the connection between 
the perturbative Callan-Symanzik $\beta$-function and the existence 
of a mass gap is addressed in Ref.~[10]). 

More importantly what the community has been neglecting is that there 
are some serious open questions regarding both the validity of 
perturbation theory in non-Abelian models and of the role of 
semiclassical approximations, which the instanton approach purports 
to be. As  stressed in several previous papers [11,12], not only 
have mathematical physicists failed to establish that ordinary 
perturbation theory does produce tha correct asymptotic
expansion of the Green's functions in powers of $1/\beta$ as 
$\beta \to \infty$ [13], but
the exactly soluble cases of spin models in 1D and gauge theories 
in 2D 
provide concrete examples in which the perturbative answers 
are (infrared)
finite yet wrong in the non-Abelian models.  Similarly, it is a 
mathematical fact that instanton computations are plagued by 
uncontrollable infrared divergences in both 2D spin modes and 
4D gauge theories [14,15].

In the last decade, several papers have tried to find support for the
orthodox point of view in either numerical studies or certain analytical 
computations, such as exact solutions in 2D or 1/N expansions. All of 
this supportive evidence is analyzed in detail in Ref.~[16]  and found 
wanting. I refer
the reader to that paper and would only like to emphasize here 
that there is complete agreement among the workers in the field 
that on thermodynamic lattices
the correlation length in spin models and the string tension in 
gauge theories increases faster than the predictions of asymptotic 
freedom. This situation is clearly in agreement with the claim 
advanced in the present paper, namely that in all $O(N)$ models 
the transition to a massless phase occurs at finite $\beta$ [17].

Assuming that a transition to a phase with algebraic decay of 
correlations does take place in all $O(N)$ models, what triggers 
it? The best way to answer is to recall the intuitively appealing 
argument advanced by Kosterlitz and Thouless many years ago [18]. 
Their idea was that topological defects are
the key to understanding phase transitions in 2D systems possessing 
continuous symmetries. Namely at low temperatures the system tries 
to minimize the energy
of the configuration. In $O(2)$ that means forming  vortices, whose energy 
nevertheless diverges like $lnL$ (L-the linear size of the lattice); 
in $O(3)$ one can form instantons, whose energy is O($L$). 
Since the entropy of a 
topological defect is of order $ln\ L$ (location of its center), 
they concluded that in $O(2)$ this type of defects are bound at 
low $1/\beta$, while they are unbound and hence producing a mass 
gap in $O(3)$. I think these considerations are 
misleading. A vortex or an instanton is a highly coherent, extended 
structure. Now one way to state the Mermin-Wagner theorem is to say 
that in the infinite
volume limit, in 2D, spin waves (configurations in which the 
gradient is 
$O(1/\sqrt\beta)$, but otherwise unconstrained) overwhelm the 
ordered configuration. If the spin waves can destroy l.r.o., 
why would they not also overwhelm the instantons with their 
tremendous entropy? In fact I would say that
the infrared divergences plaguing instanton computations are the 
sign that such an effect does happen, just as the infrared 
divergences encountered in perturbing around the ordered state 
are a manifestation of the Mermin-Wagner 
theorem. I would therefore guess that, as advocated in a previous 
paper [19], the low temperature phase is dominated by spin waves 
in all $O(N)$ $N\geq 2$ models. 
This should be true in all dimensions $D$, the special feature of 
$D=2$ being that there the invariant Green's functions of the massless 
Gaussian decay algebraically, instead of exhibiting l.r.o.
 
I would also conjecture that the non-Abelian 2D sigma models 
and 4D gauge 
theories may have a complicated behavior at low temperatures. 
Namely, at intermediate distances $(O(exp(c\beta))$ Green's functions 
may behave approximately as it was expected - exponential decay. 
However at truly asymptotic distances they 
decay algebraically, as indicated by the arguments presented in 
this paper. The only difference between Abelian and non-Abelian 
models may be that in the former,
this intermediate distance behavior is missing. (This difference may explain why non-Abelian models do not fit naturally into the conformal field theory classification.) There is a classic 
example of a physical system exhibiting such an intermediate behavior: 
the time evolution
of the probability to find a quantum particle trapped inside a potential 
well. Everybody knows about the life-time of an alpha-particle. What 
fewer people know is that one can prove rigorously that if the state 
decays at all, then 
the asymptotic behavior must be algebraic [20]. To my knowledge, all 
experimental attempts to detect the deviations from the exponential law 
in metastable
states have failed. Probably the reason is that by the time the algebraic
decay becomes important, the probability is too small to measure. 
Assuming my conjecture is correct, for $x>>1$ the 2-point function 
would behave as